\documentclass[runningheads]{llncs}
\usepackage{xcolor}
\usepackage[utf8]{inputenc}
\usepackage{hyperref}
\usepackage{listings}
\usepackage{makecell}
\usepackage{xspace}

\definecolor{cerulean}{rgb}{0, 0.4824, 0.6549}

\newcommand{\desc}[1]{{}}
\newcommand{\Soda}{\textsc{Soda}\xspace}

\hypersetup{colorlinks=true, unicode=true, linkcolor=[rgb]{0.10,0.05,0.67}, citecolor=[rgb]{0.10,0.05,0.67}, filecolor=[rgb]{0.10,0.05,0.67}, urlcolor=[rgb]{0.10,0.05,0.67}}

\begin{document}
    \lstdefinelanguage{Soda}{
    keywords=[1]{lambda, any, def, if, then, else, match, case, class, extends, abstract, end, this, subtype, supertype, false, true, not, and, or, package, import, directive, @new, @tailrec, @override},
    keywordstyle=[1]\color{teal},
    keywords=[2]{by, cases, constructor, exact, forall, fun, funext, have, induction, intro, notation, rewrite, rfl, rw, simp, theorem, with},
    keywordstyle=[2]\color{purple},
    sensitive=true,
    morecomment=[s]{/*}{*/},
    morestring=[b]"
}

\lstset{frame=tb,
    language=Soda,
    aboveskip=3mm,
    belowskip=3mm,
    showstringspaces=false,
    columns=flexible,
    basicstyle={\small\ttfamily},
    numbers=none,
    numberstyle=\tiny\color{gray},
    keywordstyle=\color{blue},
    commentstyle=\color{gray},
    stringstyle=\color{teal},
    breaklines=true,
    breakatwhitespace=true,
    tabsize=3
}

\newcommand{\srccode}[1]{\texttt{{#1}}}
\newcommand{\basicType}[1]{{\color{black}\srccode{#1}}\xspace}
\newcommand{\annotation}[1]{{\color{brown}\srccode{#1}}\xspace}

%
%

\newcommand{\reservedWordSoda}[1]{{\ensuremath{\mbox{\textbf{\srccode{#1}}}}}\xspace}
\newcommand{\sodaType}[1]{\basicType{{#1}}}

\newcommand{\sodadefequal}{\srccode{=}\xspace}
\newcommand{\sodacolon}{\srccode{:}\xspace}
\newcommand{\sodaarrow}{{-\textgreater}\xspace}
\newcommand{\sodaimplies}{{==\textgreater}\xspace}
\newcommand{\sodadefparam}{\srccode{:=}\xspace}
\newcommand{\sodalambdaarrow}{\ensuremath{\longrightarrow}\xspace}
\newcommand{\sodacasearrow}{\ensuremath{\Longrightarrow}\xspace}

\newcommand{\sodalambda}{\reservedWordSoda{lambda}}
\newcommand{\sodaany}{\reservedWordSoda{any}}

\newcommand{\sodadef}{\reservedWordSoda{def}}
\newcommand{\sodaif}{\reservedWordSoda{if}}
\newcommand{\sodathen}{\reservedWordSoda{then}}
\newcommand{\sodaelse}{\reservedWordSoda{else}}

\newcommand{\sodamatch}{\reservedWordSoda{match}}
\newcommand{\sodacase}{\reservedWordSoda{case}}

\newcommand{\sodaclass}{\reservedWordSoda{class}}
\newcommand{\sodaextends}{\reservedWordSoda{extends}}
\newcommand{\sodaabstract}{\reservedWordSoda{abstract}}
\newcommand{\sodaendclass}{\reservedWordSoda{end}}
\newcommand{\sodathis}{\reservedWordSoda{this}}
\newcommand{\sodasubtype}{\reservedWordSoda{subtype}}
\newcommand{\sodasupertype}{\reservedWordSoda{supertype}}

\newcommand{\sodafalse}{\reservedWordSoda{false}}
\newcommand{\sodatrue}{\reservedWordSoda{true}}
\newcommand{\sodanot}{\reservedWordSoda{not}}
\newcommand{\sodaand}{\reservedWordSoda{and}}
\newcommand{\sodaor}{\reservedWordSoda{or}}

\newcommand{\sodapackage}{\reservedWordSoda{package}}
\newcommand{\sodaimport}{\reservedWordSoda{import}}

\newcommand{\sodadirective}{\reservedWordSoda{directive}}

\newcommand{\sodanew}{\annotation{@new}}
\newcommand{\sodatailrec}{\annotation{@tailrec}}
\newcommand{\sodaoverride}{\annotation{@override}}

\newcommand{\sodaequalsSign}{\reservedWordSoda{==}}

\newcommand{\sodalessthancolon}{\reservedWordSoda{<:}}
\newcommand{\sodagreaterthancolon}{\reservedWordSoda{>:}}

    \sloppy
    \title{Can Proof Assistants Verify Multi-Agent Systems?}
    \author{Julian~Alfredo~Mendez\orcidID{0000-0002-7383-0529}, Timotheus~Kampik\orcidID{0000-0002-6458-2252}}

    \institute{
        Umeå University, Sweden \\
        \email{\{julian.mendez,tkampik\}@cs.umu.se}
    }
    \authorrunning{J. A. Mendez, T. Kampik}
    \maketitle
    \begin{abstract}
        This paper presents the \Soda language for verifying multi-agent systems. \Soda is a high-level functional and object-oriented language that supports the compilation of its code not only to Scala, a strongly statically typed high-level programming language, but also to Lean, a proof assistant and programming language. Given these capabilities, \Soda can implement multi-agent systems, or parts thereof, that can then be integrated into a mainstream software ecosystem on the one hand and formally verified with state-of-the-art tools on the other hand.
        We provide a brief and informal introduction to \Soda and the aforementioned interoperability capabilities, as well as a simple demonstration of how interaction protocols can be designed and verified with \Soda. In the course of the demonstration, we highlight challenges with respect to real-world applicability.
    \end{abstract}


    \section{Introduction}
    \label{sec:intro}

    \desc{What are MAS ?} Multi-agent systems (MAS) are systems composed of multiple interacting intelligent agents, working towards joint or conflicting goals. MAS have emerged as an approach for modeling and implementing complex distributed systems, because MAS-based systems can solve problems that are very difficult or impossible to solve using monolithic architectures.
    Due to their dynamic and distributed nature, MAS are notoriously difficult to test.
    At the same time, ensuring desirable behavior in MAS components is crucial to verify that the autonomy that agents are granted does not lead to safety or compliance issues.
    Consequently, testing and verification of MAS has emerged as a substantial branch of MAS engineering research.
    Here, most research focuses either on finite state model checking~\cite{DBLP:conf/kbse/BordiniDFF08} or--presumably less frequently--on testing in the software engineering sense (as in \emph{test-driven development})~\cite{DBLP:conf/atal/AmaralHK23}.
    In this paper, we take a novel path and present a prototypical approach where parts of a MAS are implemented in \Soda, a high-level programming language that can be compiled to Scala for integration with a mainstream programming ecosystem, as well as to Lean, a proof assistant and programming language that applies calculus of constructions~\cite{10.1007/978-3-319-21401-6_26}.
    The formal verification of the program code can then be executed in Lean.

    To illustrate the approach and demonstrate its feasibility, we provide a simple running example of interaction protocol verification with \Soda and Lean, which we introduce conceptually below.
    In our example, we want to implement an interaction protocol for a platform on which multiple agents buy and sell items. We assume that the transactions are safe and conducted by a mediator that is trusted by all involved agents and can transfer possession of items and money between the agents.
    In the following, we provide an example interaction sequence that can serve as an archetype for the protocol that our MAS should enact.

    \begin{enumerate}
        \item [1.] A seller $C$ notifies the mediator $A$ to advertise a certain item $R$ at a price $P$.
        \item [2.] The mediator $A$ informs all agents that the item is available for sale.
        \item [3.] While no buyer has shown interest, $C$ can remove the ad.
        \item [4.] If a buyer $B$ wants to buy $R$ and has money $P$ to pay for it, the buyer notifies $A$.
        \item [5.] When $A$ is notified of the purchase, it removes the ad, and it transactionally transfers the item $R$ from $C$ to $B$, and the amount of money $P$ from $B$ to $C$.
    \end{enumerate}

    In our approach, we use lists to store items. As a proof of concept, we prove a technical but important property:

    \textbf{Property.} \emph{Changing an item in the list does not change the size of the list}. \\

    This property is important, but it is usually taken for granted. Proving these kinds of properties in a proof assistant like Lean requires substantial effort because lemmas and tactics require maturity to be employed. All cases need to be covered, and the resulting object after the execution must always be defined. In the particular case of our functions, we work with immutable lists. Our lists are thread-safe, which allows them to run in parallel execution, and follow the standard construction in functional languages. They are defined as being either an empty list or an element prepended to a list. This makes them memory-efficient because replacing an arbitrary position requires reconstructing the list until the given position.

    As we can see, the property we prove looks simplistic from an intuitive software engineering perspective.
    Still, we chose to prove this property as a minimal demonstrating example, due to the following reasons: \emph{i)} even such simple facts are somewhat effortful to prove with proof assistants such as Lean; \emph{ii)} our example is sufficient for highlighting practical challenges, while also serving as a proof-of-concept with respect to purely technical feasibility.

    In the remainder of this paper, we give a brief overview of research on testing and verifying MAS (Section~\ref{sec:background-verification}) to then introduce \Soda and its interoperability capabilities, both with the Scala/Java ecosystem and with the Lean proof assistant (Section~\ref{sec:proof-automation}). We then continue our running example to demonstrate how agent interaction protocols can be specified and verified with \Soda (Section~\ref{sec:soda-for-mas}). Subsequently, we discuss in detail the formal proofs (Section~\ref{sec:proofs}). We show with practical experiments that the example is efficient (Section~\ref{sec:experiments}). After that, we discuss our work and its limitations, and conclude with a future outlook (Section~\ref{sec:conclusion}).


    \section{Why Formal Verification for MAS?}
    \label{sec:background-verification}
    \desc{What is software verification} Software verification ensures that specific components meet specification requirements. We can compare software verification with software validation, since both are software evaluation processes. The former focuses on determining whether a software artifact meets the conditions with respect to its broader purpose, imposed at the beginning of a development phase, whereas the latter determines whether the requirements are satisfied at the end of a development phase.
    The verification of agents and MAS has been subject to extensive study over the years.
    Here, \emph{verification} can either be seen from an applied software engineering perspective, i.e., \emph{testing} without ``hard'' guarantees, or as the formal analysis (\emph{formal verification}) of MAS or components thereof.

    Jason~\cite{Bordini-2007-Jason} and SARL~\cite{Rodriguez-2014-Sarl} are specialized multi-agent programming languages, with Jason focusing on logical reasoning and SARL on dynamic, distributed environments.
    Jason is an interpreter for an extended version of AgentSpeak(L)~\cite{Rao-1996-AgentSpeak}, a BDI (Belief-Desire-Intention) agent-oriented logic programming language. Jason provides a platform for developing multi-agent systems with customizable features, especially for environments requiring logical reasoning and decision-making.
    SARL is a general-purpose agent-oriented programming language designed to handle concurrency, distribution, interaction, decentralization, reactivity, autonomy, and dynamic reconfiguration. It integrates well with the Janus\cite{Galland-2010-Janus} platform for distributed multi-agent systems.

    Although agent testing is certainly a relevant research domain, e.g., in the context of technologies for agent-oriented programming~\cite{DBLP:conf/atal/AmaralHK23,amaraldemoaamas}, a well-informed assumption is that for MAS, testing is insufficient because of agents' and MAS' highly dynamic behavior~\cite{Winikoff2015}, in particular if contrasted with mainstream software components in server-client architectures.
    Because mainstream software testing approaches are often deemed insufficient for engineering MAS, researchers have devised a range of formal approaches to verify agents or other components of MAS.
    These approaches typically rely on model checking techniques~\cite{DBLP:conf/kbse/BordiniDFF08,modelCheckingMas,Lomuscio-2017-MCMAS}.
    Still, due to their dynamic nature and the resulting combinatorial explosion of the number of models~\cite{Winikoff2015}, applying model checking to MAS can be similarly intractable as attempting to achieve substantial coverage in traditional tests.
    Also, model-checking approaches are typically limited to particular abstractions and modalities; for example, TLA+ prominently focuses on temporal reasoning~\cite{DBLP:books/aw/Lamport2002}.
    In contrast, proof assistants such as Coq~\cite{barras1997coq}, Isabelle~\cite{paulson1994isabelle}, and Lean~\cite{de2015lean} allow for more generic formal analyses, and thus they can potentially enable more flexible verification of agents and MAS.
    However, these tools are not straightforwardly applicable to the formal analysis of program code written in languages that engineers use to specify behavior at the ``business logic''(or \emph{knowledge}) level.
    This may explain why proof assistants have not yet been comprehensively studied in the context of MAS verification, at least (to the best of our knowledge) not as tools that can directly analyze executable specifications of MAS components.

    The work we present in this paper is a step towards overcoming this obstacle, by bridging the gap between mainstream software ecosystems in which MAS can be engineered and tools for proof automation, demonstrating basic feasibility by implementing a simple example.


    \section{Designing MAS with \Soda}
    \label{sec:proof-automation}

    Before we demonstrate \Soda's verification capabilities, let us provide a brief overview of its syntax (Subsection~\ref{subsec:syntax}) and the technologies the language is built upon (Subsection~\ref{subsec:tech}).

    \subsection{\Soda syntax}
    \label{subsec:syntax}

    \desc{What is \Soda.} \Soda~\cite{Mendez-2023-Soda}\footnote{For a more detailed technical report that describes the \Soda language, see~\cite{Mendez-2023-Soda}.} is a statically typed functional language with object-oriented notation. Its name is an acronym for \emph{Symbolic Objective Descriptive Analysis}, and it was designed to be descriptive by being declarative and easy to read.
    It combines the rigor of purely functional languages like Lean, with the modern technology integration of languages like Scala, which in turn is a bridge to the Java Virtual Machine (JVM) ecosystem.
    \Soda provides not only the expressive power for the type of transactions we have in the example, but also the elements to include Lean proofs for the definitions. It also allows for interconnection with complex packages provided by JVM libraries and frameworks, and it has a simple and direct way to write functions, similar to mainstream languages like Scala or Python.

    \desc{What are the main constructs of \Soda} \Soda follows the functional style of ML\cite{Milner-1978-Theory}\footnote{ML stands for \emph{Meta Language}, but the acronym is rarely spelled-out.}, and it has a small set of constructs. Most of them can be placed in one of the following categories: function-related constructs and class-related constructs. It uses the basic types provided by Scala, such as Boolean, Int, Float, and String.

    \desc{Function-related constructs} The function-related constructs are meant to define functions. It is possible to say that a variable $x$ is of a certain type $A$ with $x : A$. The functions are then defined in the usual way; e.g., $f (x : Int) : Int = x + 1$ defines the function $x + 1$. Lambda expressions are also written as expected: the function above can be defined using the lambda expression $f : Int = \sodalambda \ x \sodalambdaarrow x + 1$. The construct \sodaif-\sodathen-\sodaelse allows for the definition of piecewise functions. As in other ML-styled functional languages, the construct \sodamatch-\sodacase-\sodacasearrow is used for pattern matching. The following function returns the maximum of two integers:
    $$max \ (x : Int) \ (y : Int) : Int = \sodaif\ x > y \ \sodathen\ x \ \sodaelse\ y$$

    \desc{Class-related constructs} The class-related constructs are meant to define new types, to group functions in classes, and to group classes in packages. The construct \sodaclass-\sodaendclass defines a class from the beginning to the end. It groups functions in a module, and it defines a type, a default constructor, and a default equality function. A class can extend other classes with the construct \sodaextends. If a class needs parameters to instantiate an object, these parameters can be defined with the construct \sodaabstract. A class can depend on parameterized types, which are parameters belonging to \texttt{Type} inside square brackets (\texttt{[ ]}). Classes are grouped in packages with the \sodapackage reserved word, and \sodaimport is used to import classes from different packages.

    \subsection{Technology behind \Soda}
    \label{subsec:tech}

    \desc{\Soda is the intersection of Scala 3 and Lean 4}
    In a figurative way, we can say that the \Soda syntax lies in the intersection of Scala version 3 syntax and Lean version 4 syntax\footnote{Scala 3 and Lean 4 have improved their syntax compared to Scala 2 and Lean 3 respectively.}, and they all fall under the umbrella of ML-styled programming languages.
    \Soda code needs to be compiled in order to be executed. The main compiler converts the \Soda source code into Scala source code. Since it is a source-to-source compiler of similar source code, this compiler is a \emph{transpiler}, and we also call it a \emph{translator}.
    We claim that the translation into Scala is sound, in the sense that all correct code in \Soda is translated into correct code in Scala. It is also possible to import Java Virtual Machine (JVM) libraries into \Soda, using \sodaimport, but it requires a manual check from the developer to ensure that the imported libraries keep the code purely functional.

    \desc{Translation to Lean} \Soda can be partially translated into Lean. This means that it is possible to prove properties of \Soda code in Lean. The construct \sodadirective can include embedded pieces of code of a specific target language. For example, a proof would be considered in a translation to Lean, but it would be ignored in a translation to Scala. The proof has to be actually written in Lean, based on definitions provided in \Soda.
    \desc{Source code} \Soda's source code is available for download at \url{https://github.com/julianmendez/soda}.


    \section{\Soda for MAS Verification}
    \label{sec:soda-for-mas}

    \desc{
        \begin{itemize}
            \item How is the example designed?
            \item How is the example implemented?
            \item Show how Lean can be used
        \end{itemize}
    }

    \desc{\Soda as a language for specification of MAS} \Soda provides a small set of constructs to create code that can be executed in Scala 3 and proven in Lean 4. On the one hand, Lean is not only a powerful programming language, but also a proof assistant. Its mathematical library, mathlib\footnote{\url{https://github.com/leanprover-community/mathlib4}},  contains a significant number of mathematical theorems and their proofs, which can serve as a template to prove properties of agents. On the other hand, Scala is a very flexible language, with an efficient implementation connected to the JVM libraries, which allows for multi-platform execution. It is important to note that Lean proofs cannot cover the use of JVM libraries, since these are not specified for Lean.

    \desc{How is the example designed} We have specified the example of Section~\ref{sec:intro} in \Soda. We describe the basic types, available functions, and some properties that the system should hold. Although the example is intended to be a proof of concept, it already contains the main components for a full-scale example.
    \desc{Basic types} All agents are identified with a unique number, which is associated with an account number controlled by the mediator. Thus, the mediator can transfer money between the accounts of the participants. The mediator can also transfer items, which are also identified with unique numbers. The items have properties like their owner and price.
    \desc{Available functions} The available functions are intended to perform the main operation between the agents. There are functions to advertise an item, to change its price, and to sell it to another agent. Since \Soda is purely functional, the functions do not change the state of a market, but create new instances of the market. The actual communication between the agents is done through a library used in the Scala translation.

    \desc{Invariants} In order to enable the verification of the desired behavior, we make use of \emph{invariants}. Invariants are logical assertions that are always valid during execution. They are helpful resources to prove that a system behaves according to predefined expectations. As Lean can be used to prove theorems, the Lean translation can serve to prove the preservation of invariants.
    The two invariants which we are interested in are the following:
    \begin{itemize}
        \item No item is created or destructed after modifying the state of an item;
        \item No account is created or destructed after modifying the state of an item.
    \end{itemize}

    \desc{Choice for representation} The choice for representation is not trivial. Different ways of representing the market lead to a number of derived challenges. Sometimes, more efficient performance is achieved by adding some redundancy, but it makes it more difficult to prove invariants. Similarly, the definitions of the functions require consideration, as shown in Figure~\ref{listing:length-def-fl} where an intuitive recursive definition of the length of a list (\texttt{length\_def}) should be equivalent to a tail recursive definition of the length of a list (\texttt{length\_fl}). We prove that both definitions are equivalent, allowing us to use the intuitive definition for the proofs and the tail recursive definition for the execution.

    \begin{figure}[h!]

        \begin{lstlisting}[label={lst:tailrec-def-fl}]
  _tailrec_foldl [A : Type] [B : Type] (list : List [A] ) (current : B)
      (next : B -> A -> B) : B =
    match list
      case Nil ==> current
      case (head) :: (tail) ==>
        _tailrec_foldl [A] [B] (tail) (next (current) (head) ) (next)

  foldl [A : Type] [B : Type] (list : List [A] ) (initial : B)
      (next : B -> A -> B) : B =
    _tailrec_foldl [A] [B] (list) (initial) (next)

  length_fl [A : Type] (list : List [A] ) : Nat =
    _tailrec_foldl [A] [Nat] (list) (0) (
      lambda (accum : Nat) --> lambda (_elem : A) --> accum + 1)

  _tailrec_length [A : Type] (list : List [A] ) (accum : Nat) : Nat =
    match list
      case Nil ==> accum
      case (_head) :: (tail) ==>
        _tailrec_length [A] (tail) (accum + 1)

  length_tr [A : Type] (list : List [A] ) : Nat =
    _tailrec_length [A] (list) (0)

  length_def [A : Type] (list : List [A] ) : Nat =
    match list
      case Nil ==> 0
      case (_head) :: (tail) ==> length_def [A] (tail) + 1
        \end{lstlisting}

        \caption{The function \texttt{\_tailrec\_foldl} is a tail-recursive auxiliary function of \texttt{foldl}, which is a left fold (\emph{foldl}) function for parameterized types.
        \texttt{length\_fl}, \texttt{length\_tr}, and \texttt{length\_def} syntactically different but semantically equivalent ways of defining the length of a list of an arbitrary type $A$ in \Soda. The tail-recursive definitions (\texttt{length\_fl}, \texttt{length\_tr}) outperform the naive definition (\texttt{length\_def}).}
        \label{listing:length-def-fl}

    \end{figure}

    \subsection{Package Structure}
    \label{subsec:package-structure}

    \desc{About the implementation and its design} The project is available at \url{https://github.com/julianmendez/market} (open-source). It is designed to be compiled using sbt \footnote{\url{https://www.scala-sbt.org}} to be executed in the JVM environment. A Bash script uses the \Soda binary to translate the \Soda files, including those proofs that can be verified with Lean. The project file structure follows the standard sbt/Maven layout. The project contains three packages:
    \begin{itemize}
        \item \texttt{core}, containing the core data types, the market, its functionality and the proofs;
        \item \texttt{parser}, containing a YAML parser; and
        \item \texttt{main}, containing the entry point class, which supports the console execution.
    \end{itemize}

    \desc{The main core} Firstly, the main core is a package that contains the classes to model the market. The main operations are:
    \begin{itemize}
        \item \textbf{deposit} $user$ $amount$ : works as an update-insert (upsert) function to declare a user and to put money in the linked account.
        \item \textbf{assign} $item$ $user$ : works as an upsert function to declare an item and to establish its owner.
        \item \textbf{price} $item$ $amount$ : has the double functionality of assigning the price of an item for sale, if the value is positive, or hiding it, if the value is zero.
        \item \textbf{sell} $item$ $user$ : transactionally transfers the possession of an item to a specified user, and money equivalent for the price from the new user's to the old user's account.
    \end{itemize}

    \desc{YAML parser} Secondly, the project has a package to read YAML files. A list of operations simulates the interaction that occurs in a MAS scenario. The parser creates a list of operations, which can later be executed on an instance of market.

    \desc{The entrypoint package} The last relevant package is the \textit{main} package, which contains the entry point and is intended to receive and parse parameters from the console.

    The parser package and the core packages contain their counterpart testing classes. The parser test package has two layers of parsing: the syntactic layer, which is implemented using SnakeYAML\footnote{\url{https://bitbucket.org/asomov/snakeyaml-engine}}, a Java YAML parser, and the semantic layer, where the operation objects are instantiated.
    The core test package instantiates a market and applies the parsed operations to it.

    Before we start modeling our \emph{market} example, we need to introduce two fundamental preliminaries that are crucial to enable formal verification: natural number support (Subsection~\ref{subsec:natural-numbers}) and list wrappers (Subsection~\ref{subsec:list-wrappers}).

    \subsection{Preliminary 1: Support for Natural Numbers}
    \label{subsec:natural-numbers}

    One of the challenges that we need to face is how we handle natural numbers in \Soda. We use the standard notation in the literature where natural numbers are used as a synonym of non-negative integers, i.e. including the number 0. While Lean has natural numbers as part of its core (\texttt{Nat}), Scala does not have natural numbers as a type. In Soda, we implement the natural numbers based on a type \texttt{Nat}, and two constructors: \texttt{0} and \texttt{Succ\_}. We define its translation as shown in Figure~\ref{listing:nat-def}.

    \begin{figure}[h!]

        \begin{lstlisting}[label={lst:nat-def}]
directive scala
type Nat = Int
object Succ_ {
  def apply (n : Int) : Int = n + 1
  def unapply (n : Int) : Option [Int] =
    if (n <= 0) None else Some (n - 1)
}

directive lean
notation "Succ_" => Nat.succ
        \end{lstlisting}

        \caption{This table shows the \Soda definition of \texttt{Nat}, to be translated to Scala and to Lean.}
        \label{listing:nat-def}

    \end{figure}

    \noindent A good definition of natural numbers is crucial in order to accomplish two things: proofs and calculations. A recursive definition is needed to provide proofs in Lean, especially those involving recursive definitions such as the length of a list. However, some calculations may need to be time efficient, even with larger natural numbers.

    \desc{monus1} In functional languages with ML syntax, the deconstruction is often done with pattern matching. While the Lean syntax allows us to use this type of syntax, in Scala this does not work off-the-shelf. The main issue is that our natural numbers in Scala are in fact \texttt{Int} objects, which makes addition of two numbers very efficient, and \texttt{Succ\_} is a construction external to the objects. For that reason, we define the function $monus1$, which subtracts 1 from a given positive number until this number reaches 0. This is defined in Figure~\ref{listing:monus1-def}. The function $monus1$ is based on the function $monus$, denoted by $\mathop {\dot -}$, which handles subtraction in natural numbers. For two natural numbers $a$ and $b$, it is defined as: $a \mathop {\dot -} b = \max (a - b , 0)$.

    \begin{figure}[h!]

        \begin{lstlisting}[label={lst:monus1-def}]
  monus1 (index : Nat) : Nat =
    match index
      case Succ_ (k) ==> k
      case _otherwise ==> 0
        \end{lstlisting}

        \caption{This table shows the definition of \emph{monus1}, which for a positive number, it returns the previous number, and for other numbers, it returns 0.}
        \label{listing:monus1-def}

    \end{figure}

    \subsection{Preliminary 2: List Wrappers}
    \label{subsec:list-wrappers}

    After having defined \texttt{Nat}, we define a type we name \texttt{ListWrapper}. This is actually not a list itself, but a toolbox that works as a wrapper for lists, containing methods and their proofs. Let us see how this type is constructed.
    $Nil$ is the constructor for an empty list, $::$ is the list constructor, and $Nat$ is a type for non-negative integers.

    \desc{The role of \emph{foldl}.}
    The function \emph{foldl} (from \emph{fold left}), as it is called in other functional programming languages like Haskell, applies a combining function to an iterable structure, a finite list in our case, and starts with an initial value.
    One relevant property of \emph{foldl} is the possibility of implementing it as a tail-recursive function, which in turn can be optimized by a virtual machine or a compiler as a cycle. In other words, \emph{foldl} is a purely functional approach to execute certain types of cycles. In addition to its efficient execution, \emph{foldl} ensures termination on finite structures, like the list structure.
    In Figure~\ref{listing:length-def-fl}, we can see \texttt{foldl}, which is an implementation of \emph{foldl} that uses an auxiliary tail recursive function called \texttt{\_tailrec\_foldl}.

    \noindent Although \emph{foldl} is a useful resource for efficient and well-founded recursion, it becomes more complex when it is involved in proofs. Because of that, in our proofs, we provide (traditionally) recursive definitions, in addition to the tail-recursive \emph{foldl} definitions.

    \desc{Accessors} \Soda lists allow for access to any arbitrary position. In Scala, accessing at position $i$ of a list $a$ of type $T$ would be notated $a(i)$, which is already a value of type $T$. However, when we need to access an element inside a proof, we need to ensure that $i$ is in range, and thus that $a(i)$ is well defined. We present this in Figure~\ref{listing:get-def}, where we show how we access an element of a list.
    The function \texttt{get} visits all the elements of a list until it reaches the specified position. If that happens, it returns the content at that position (\texttt{Some (elem)}), otherwise, it returns \texttt{None}.

    \begin{figure}[h!]

        \begin{lstlisting}[label={lst:get-def}]

  _tailrec_get_def [A : Type] (list : List [A] ) (index : Nat) (current : Nat) : Option [A] =
    match list
      case Nil ==> None
      case (head) :: (tail) ==>
        if current == index
        then Some (head)
        else _tailrec_get_def [A] (tail) (index) (Succ_ (current) )

  get [A : Type] (list : List [A] ) (index : Nat) : Option [A] =
    _tailrec_get_def [A] (list) (index) (0)

        \end{lstlisting}

        \caption{The function \texttt{get} retrieves an element safely, without throwing an exception if the index is out of range. This is necessary to ensure that the retrieved element is well defined.}
        \label{listing:get-def}

    \end{figure}

    \desc{Mutability} Since \Soda lists are immutable, modifying the state of a market requires creating new lists. This can be written in Scala for a mutable structure as $a(i) = e$, which updates the content of a list $a$ at position $i$, by assigning it the value $e$.
    In Figure~\ref{listing:set-def}, we show the definition of \texttt{set\_def}, and in Section~\ref{sec:proofs}, we prove that it preserves the length of the list, i.e. it does not create or remove elements.

    \begin{figure}[h!]

        \begin{lstlisting}[label={lst:set-def}]
  set_def [A : Type] (list : List [A] ) (index : Nat) (element : A) : List [A] =
    match list
      case Nil ==> Nil
      case (head) :: (tail) ==>
        if index == 0
        then (element) :: (tail)
        else (head) :: (set_def [A] (tail) (monus1 (index) ) (element) )

        \end{lstlisting}

        \caption{The function \texttt{set\_def} is a tail recursive function that creates a new list with an element updated at a given index.}
        \label{listing:set-def}

    \end{figure}

    \subsection{Modeling the Market}
    \label{subsec:market}
    As we can see in Figure~\ref{listing:market-struc}, money is modeled as an integer. An item is modeled as a pair $\langle$owner, price$\rangle$. A market is a pair containing the owner's money and the items. Conventionally, we say that an item is advertised if and only if its price is greater than 0. In the context of this example, money is modeled as a natural number (\texttt{Nat}), but in reality, we could allow users to have debt and use an integer instead.

    \begin{figure}[h!]

        \begin{lstlisting}[label={lst:market-struc}]
class Money = Int

class Item

  abstract
    owner : Nat
    price : Money

end

class Market

  abstract
    accounts : List [Money]
    items : List [Item]

end
        \end{lstlisting}

        \caption{A market is a structure composed of smaller structures.}
        \label{listing:market-struc}

    \end{figure}

    \desc{Why a module} The methods for manipulating a market or an item are separated in a module called \texttt{MarketMod}. This design style differs from the traditional object-oriented approach, where the methods pertaining to the market should be in the market class itself. We chose this design as a compromise in the integration of \Soda, Scala, and Lean. While Scala uses a traditional object notation as in other object-oriented programming languages, Lean uses syntactic sugar to access functions inside modules, as in other functional programming languages. Since the \Soda translators do not support the notational translation between the two paradigms, we place the functions in modules, and use classes as structures.

    \section{Formal Proofs}
    \label{sec:proofs}

    \desc{What we would like to compute} As we discuss above, the virtue of \Soda resides in connecting efficient implementations and formal proofs in the same language. We decided to show a simple property: ``Changing an element in a list does not modify its length''. This property allows ensuring that a transaction does not change the length of a list. This could be reformulated as: ``Creating a list with \texttt{set\_def} returns a list of the same length as the original''.

    \subsection{Defining the Length of a List}

    To prove the goal, we use the definition of length given in Figure~\ref{listing:length-def-fl}. Although this definition looks clear, it is not tail recursive, and for that reason it could be considered a naive definition of the length of a list. For example, in an average computer configuration, this function cannot compute the length of a list of 10,000 elements, since it would produce a stack overflow during execution. However, this definition greatly simplifies the proofs.
    A more applicable function to compute the length of a list is \texttt{length\_tr}, which uses tail recursion. Its implementation is shown in Figure~\ref{listing:length-def-fl}.

    Figure~\ref{listing:length-equiv} shows Theorem~\texttt{len\_tr\_eq\_len\_def}, which states that the tail recursive function is equivalent to the naive definition. The proof uses Lemma~\texttt{len\_tr\_accum}, which helps in the interpretation of the extra parameter of \texttt{\_tailrec\_length}.

    \desc{Proof of the lemma} Although fully understanding the proof requires understanding of Lean 4, we sketch how the proof is structured. Let us focus on Lemma~\texttt{len\_tr\_accum} in Figure~\ref{listing:length-equiv}, which proves that the $accum$ parameter accumulates the computed length. To achieve this, we apply an inductive strategy to the list (\texttt{induction list with}). If the list is empty (\texttt{nil}), it rewrites the definitions. If the list is not empty (\texttt{cons head tail ih}), it applies some definitions and uses the inductive hypothesis for the given parameters. Notice that the property has a value $accum$, which we can instantiate to different values to use the induction hypothesis.

    \desc{Proof of the theorem} This theorem is in fact shorter than the lemma, and it is also proven by applying induction. Since the equivalence does not have parameters, we first add them (\texttt{funext A list}). As before, the induction has two cases. The \texttt{nil} case is a direct rewriting of the definition. For the inductive step (\texttt{cons head tail ih}), we use the lemma and then the induction hypothesis.

    \begin{figure}[h!]

        \begin{lstlisting}[label={lst:length-equiv}]
  directive lean
  theorem
    len_tr_accum (A : Type) (list : List (A) )
      : forall (accum : Nat) ,
        _tailrec_length (A) (list) (accum)  = _tailrec_length (A) (list) (0) + accum := by
      induction list with
      | nil =>
        intro n
        rewrite [_tailrec_length, _tailrec_length, Nat.zero_add]
        rfl
      | cons head tail ih =>
        intro n
        rewrite [_tailrec_length, _tailrec_length]
        rewrite [ih (1)]
        rewrite [ih (n + 1)]
        rewrite [Nat.add_assoc]
        rewrite [Nat.add_comm 1]
        rfl

  directive lean
  theorem
    len_tr_eq_len_def
      : length_tr = length_def := by
    funext A list
    rewrite [length_tr]
    induction list with
    | nil =>
      rewrite [_tailrec_length, length_def]
      rfl
    | cons head tail ih =>
      rewrite [_tailrec_length, len_tr_accum]
      rewrite [ih]
      rewrite [length_def]
      rfl
        \end{lstlisting}

        \caption{Proof of Theorem \texttt{len\_tr\_eq\_len\_def}: it states that \texttt{length\_tr} = \texttt{length\_def}. This means that the tail recursive definition of length of a list is equivalent to the naive definition. The proof uses Lemma \texttt{len\_tr\_accum}.}
        \label{listing:length-equiv}

    \end{figure}

    Since we have proven that both functions are equivalent, we can use the efficient function (\texttt{length\_tr}) for the definitions, and the naive function (\texttt{length\_def}) for the proofs.

    \subsection{Updating a List}

    \desc{The length of set\_def} We have auxiliary functions and lemmas to prove that \texttt{set\_def} does not change the number of elements. Figure~\ref{listing:monus1-lemma} shows a property on \texttt{monus1} stating that applying it to the successor of a number returns the original number.
    In Figure~\ref{listing:len-set-theorem}, we can see the proof that   \texttt{set\_def} preserves the length of the list, and the proof is provided by induction on the list. For the base case of the empty list (\texttt{nil}), it suffices to provide the definitions of \texttt{set\_def} and \texttt{length\_def}. For the inductive case (\texttt{cons head tail ih}), after using the definitions of \texttt{set\_def} and \texttt{length\_def}, we split the proof into two cases based on the value of the index. If it is at the beginning, (\texttt{zero}), the definition of \texttt{monus1} is used. Otherwise (\texttt{succ k}), after using the definitions of \texttt{monus1} and \texttt{length}, we employ the induction hypothesis.

    \begin{figure}[h!]

        \begin{lstlisting}[label={lst:monus1-lemma}]
  directive lean
  theorem
    monus1_succ
      : forall (index : Nat),
        monus1 (Nat.succ (index)) = index := by
    intro idx
    rewrite [monus1]
    simp
        \end{lstlisting}

        \caption{Lemma proving that $monus1$ works as the inverse of successor.}
        \label{listing:monus1-lemma}

    \end{figure}

    \begin{figure}[h!]

        \begin{lstlisting}[label={lst:len-set-theorem}]
  directive lean
  theorem
    len_set (A : Type) (list : List (A)) (element : A)
      : forall (index : Nat),
        length_def (A) (set_def (A) (list) (index) (element) ) = length_def (A) (list) := by
    induction list with
    | nil =>
      intro idx
      rewrite [set_def, length_def]
      rfl
    | cons head tail ih =>
       intro idx
       rewrite [set_def, length_def]
       cases idx with
       | zero =>
         rewrite [monus1]
         rewrite [Nat.zero_eq]
         rfl
       | succ k =>
         rewrite [monus1]
         simp
         rewrite [length_def]
         rewrite [ih]
         rfl
        \end{lstlisting}

        \caption{Theorem and proof that proves that \texttt{set\_def} does not change the length of the provided list.}
        \label{listing:len-set-theorem}

    \end{figure}


    \section{Experiments}
    \label{sec:experiments}

    \desc{Description of the experiments} To test the efficiency of the market application, we run a series of experiments. The purpose of the experiments is to determine the viability of the example.
    As a prerequisite, we have developed a tool that generates test instances. Each instance depends on three parameters: the number of users, the number of items, and the number of transactions.

    \desc{Structure of the instances} Each test instance is a YAML file that follows some patterns. At the beginning, it contains the instructions to add the user accounts, with some amount of money. After that, it adds the items assigned to some users, with some non-zero price. Finally, it adds all sell transactions together with a price change. This price change is needed because the market puts an item as not-for-sale after each transaction is successful.

    \desc{Purpose of the experiments} We fix some values, and we assume that each user has, on average, $8$ items to sell. We conducted two types of experiments, which are shown in Table~\ref{table:fixed-transac}. On the left, we choose an arbitrary batch of 65536 ($2^{16}$) transactions and we test the system with an exponentially growing number of users and items. On the right, we fix the number of users and items as 2048 and 16384 respectively, and we test the instances with an exponentially growing number of transactions.
    The experiments were run on a Linux Ubuntu 22.04.5 LTS computer equipped with 8 Intel cores i5-8350U CPU at 1.70 GHz and 32 GiB of RAM.
    The Bash script to run the experiments is provided in the project repository.

    \begin{table}[h!t]
        \caption{The elapsed time in seconds of different synthetically generated instances.
        The columns show: the number of users, the number of items, the number of transactions, the elapsed time in seconds, and the number of transactions per second.}
        \label{table:fixed-transac}

        \begin{center}
            \begin{tabular}{ccc}
                \begin{tabular}{rrrrr}
                    \hline
                    \textbf{users} & \textbf{items} & \textbf{transac} & \textbf{time (s)} & \textbf{tr/s} \\
                    \hline
                    1              & 8              & 65536            & 1.10              & 59578         \\
                    2              & 16             & 65536            & 1.13              & 57996         \\
                    4              & 32             & 65536            & 1.13              & 57996         \\
                    8              & 64             & 65536            & 1.26              & 52013         \\
                    16             & 128            & 65536            & 1.40              & 46811         \\
                    32             & 256            & 65536            & 1.48              & 44281         \\
                    64             & 512            & 65536            & 1.84              & 35617         \\
                    128            & 1024           & 65536            & 2.32              & 28248         \\
                    256            & 2048           & 65536            & 3.55              & 18461         \\
                    512            & 4096           & 65536            & 5.79              & 11319         \\
                    1024           & 8192           & 65536            & 10.41             & 6295          \\
                    2048           & 16384          & 65536            & 19.89             & 3295          \\
                    \hline
                \end{tabular} & \ \ \ \ \  &
                \begin{tabular}{rrrrr}
                    \hline
                    \textbf{users} & \textbf{items} & \textbf{transac} & \textbf{time (s)} & \textbf{tr/s} \\
                    \hline
                    2048           & 16384          & 256              & 4.35              & 59            \\
                    2048           & 16384          & 512              & 5.05              & 101           \\
                    2048           & 16384          & 1024             & 4.95              & 207           \\
                    2048           & 16384          & 2048             & 4.87              & 421           \\
                    2048           & 16384          & 4096             & 5.28              & 776           \\
                    2048           & 16384          & 8192             & 6.36              & 1288          \\
                    2048           & 16384          & 16384            & 8.23              & 1991          \\
                    2048           & 16384          & 32768            & 13.14             & 2494          \\
                    2048           & 16384          & 65536            & 19.89             & 3295          \\
                    2048           & 16384          & 131072           & 36.67             & 3574          \\
                    2048           & 16384          & 262144           & 68.49             & 3827          \\
                    2048           & 16384          & 524288           & 126.88            & 4132          \\
                    \hline
                \end{tabular} \\
            \end{tabular}
        \end{center}

    \end{table}

    \desc{What we read from the graphs} From Table~\ref{table:fixed-transac}, we conclude that the measurements show the effect of the overhead starting time of the application, as the first rows present very similar values. Another result is that the growth in the number of users and items seems to have a linear impact on the execution time. Regarding the transactions, once the number of users and the number of items are fixed, the required time appears to be linear with respect to the number of transactions. And, most importantly, with an average of around 4000 transactions per second, we demonstrate the viability of our example: the results indicate that an application of a multi-agent environment with formally verified pieces of code can also be efficient to run in production.

    \section{Conclusion}
    \label{sec:conclusion}
    \desc{Conclusion: Formal verification for MAS is relevant} In this paper, we have demonstrated the applicability of \Soda to the design of (so far simple) MAS components that can be integrated into the Scala and Java technology ecosystem, while also supporting formal verification with the proof assistant Lean. We claim that the use of proof assistants like Lean for MAS verification is a relevant research direction, as it allows for a more abstract and flexible formal analysis of MAS.
    However, given our demonstration, it is clear that more work is required to support the verification of MAS in a meaningful manner that is useful for software engineers.
    To move towards applicability, we hope to further advance this research in the following ways: \emph{i)} by expanding proof-of-concept implementations to more complex scenarios that require the verification of additional fundamental MAS abstractions, such as agents' reasoning loops, providing reusable abstractions for verification; \emph{ii)} by providing a more detailed analysis of the advantages and disadvantages of using proof assistants instead of model checkers for MAS verification; \emph{iii)} by studying the applicability of our formal verification approach to real-world problems, such as the assurance of fairness properties in complex socio-technical systems.
    As an independent research direction, we consider it relevant to investigate whether and to what extent existing agent programming languages, such as protocol-based languages~\cite{DBLP:conf/atal/VSC23,DBLP:conf/atal/VSC23a}, can be extended or integrated to allow formal verification with proof assistants.

    \bibliographystyle{splncs04}

    \bibliography{main}

\begin{thebibliography}{10}
\providecommand{\url}[1]{\texttt{#1}}
\providecommand{\urlprefix}{URL }
\providecommand{\doi}[1]{https://doi.org/#1}

\bibitem{DBLP:conf/atal/AmaralHK23}
Amaral, C.J., H{\"{u}}bner, J.F., Kampik, T.: {TDD} for {AOP:} test-driven
  development for agent-oriented programming. In: Agmon, N., An, B., Ricci, A.,
  Yeoh, W. (eds.) Proceedings of the 2023 International Conference on
  Autonomous Agents and Multiagent Systems, {AAMAS} 2023, London, United
  Kingdom, 29 May 2023 - 2 June 2023. pp. 3038--3040. {ACM} (2023).
  \doi{10.5555/3545946.3599165},
  \url{https://dl.acm.org/doi/10.5555/3545946.3599165}

\bibitem{amaraldemoaamas}
Amaral, C.J., Kampik, T., Cranefield, S.: {A Framework for Collaborative and
  Interactive Agent-oriented Developer Operations}. In: Proceedings of the 19th
  International Conference on Autonomous Agents and MultiAgent Systems. AAMAS
  ’20, International Foundation for Autonomous Agents and Multiagent Systems,
  Richland, SC (2020)

\bibitem{barras1997coq}
Barras, B., Boutin, S., Cornes, C., Courant, J., Filliatre, J.C., Gimenez, E.,
  Herbelin, H., Huet, G., Munoz, C., Murthy, C., et~al.: {The Coq proof
  assistant reference manual: Version 6.1} (1997)

\bibitem{DBLP:conf/kbse/BordiniDFF08}
Bordini, R.H., Dennis, L.A., Farwer, B., Fisher, M.: {Automated Verification of
  Multi-Agent Programs}. In: 23rd {IEEE/ACM} International Conference on
  Automated Software Engineering {(ASE} 2008), 15-19 September 2008, L'Aquila,
  Italy. pp. 69--78. {IEEE} Computer Society (2008). \doi{10.1109/ASE.2008.17},
  \url{https://doi.org/10.1109/ASE.2008.17}

\bibitem{Bordini-2007-Jason}
Bordini, R.H., H{\"u}bner, J.F., Wooldridge, M.: {Programming multi-agent
  systems in AgentSpeak using Jason}, vol.~15. John Wiley \& Sons (2007)

\bibitem{de2015lean}
De~Moura, L., Kong, S., Avigad, J., Van~Doorn, F., von Raumer, J.: {The Lean
  theorem prover (system description)}. In: Automated Deduction-CADE-25: 25th
  International Conference on Automated Deduction, Berlin, Germany, August 1-7,
  2015, Proceedings 25. pp. 378--388. Springer (2015)

\bibitem{modelCheckingMas}
Dennis, L.A., Fisher, M., Webster, M.P., Bordini, R.H.: {Model checking agent
  programming languages}. Automated Software Engineering  \textbf{19}(1),
  5--63 (2012). \doi{10.1007/s10515-011-0088-x},
  \url{https://doi.org/10.1007/s10515-011-0088-x}

\bibitem{Galland-2010-Janus}
Galland, S., Gaud, N., Rodriguez, S., Hilaire, V.: {Janus: another yet
  general-purpose multiagent platform}. In: Proceedings of 7th Agent-Oriented
  Software Engineering Technical Forum (TFGASOSE-10) (2010)

\bibitem{DBLP:books/aw/Lamport2002}
Lamport, L.: {Specifying Systems, The {TLA+} Language and Tools for Hardware
  and Software Engineers}. Addison-Wesley (2002),
  \url{http://research.microsoft.com/users/lamport/tla/book.html}

\bibitem{Lomuscio-2017-MCMAS}
Lomuscio, A., Qu, H., Raimondi, F.: {MCMAS: an open-source model checker for
  the verification of multi-agent systems}. International Journal on Software
  Tools for Technology Transfer  \textbf{19}(1),  9--30 (2017).
  \doi{10.1007/s10009-015-0378-x},
  \url{https://doi.org/10.1007/s10009-015-0378-x}

\bibitem{Mendez-2023-Soda}
Mendez, J.A.: {Soda: An Object-Oriented Functional Language for Specifying
  Human-Centered Problems} (2023). \doi{10.48550/arXiv.2310.01961}

\bibitem{Milner-1978-Theory}
Milner, R.: {A theory of type polymorphism in programming}. Journal of Computer
  and System Sciences  \textbf{17}(3),  348--375 (1978).
  \doi{10.1016/0022-0000(78)90014-4},
  \url{https://www.sciencedirect.com/science/article/pii/0022000078900144}

\bibitem{10.1007/978-3-319-21401-6_26}
de~Moura, L., Kong, S., Avigad, J., van Doorn, F., von Raumer, J.: {The Lean
  Theorem Prover (System Description)}. In: Felty, A.P., Middeldorp, A. (eds.)
  Automated Deduction - CADE-25. pp. 378--388. Springer International
  Publishing, Cham (2015)

\bibitem{paulson1994isabelle}
Paulson, L.C.: {Isabelle: A generic theorem prover}. Springer (1994)

\bibitem{Rao-1996-AgentSpeak}
Rao, A.S.: {AgentSpeak(L): BDI agents speak out in a logical computable
  language}. In: Van~de Velde, W., Perram, J.W. (eds.) Agents Breaking Away.
  pp. 42--55. Springer Berlin Heidelberg, Berlin, Heidelberg (1996)

\bibitem{Rodriguez-2014-Sarl}
Rodriguez, S., Gaud, N., Galland, S.: {SARL: a general-purpose agent-oriented
  programming language}. In: 2014 IEEE/WIC/ACM International Joint Conferences
  on Web Intelligence (WI) and Intelligent Agent Technologies (IAT). vol.~3,
  pp. 103--110. IEEE (2014)

\bibitem{DBLP:conf/atal/VSC23}
V., S.H.C., Singh, M.P., Chopra, A.K.: {Kiko: Programming Agents to Enact
  Interaction Models}. In: Agmon, N., An, B., Ricci, A., Yeoh, W. (eds.)
  Proceedings of the 2023 International Conference on Autonomous Agents and
  Multiagent Systems, {AAMAS} 2023, London, United Kingdom, 29 May 2023 - 2
  June 2023. pp. 1154--1163. {ACM} (2023). \doi{10.5555/3545946.3598758},
  \url{https://dl.acm.org/doi/10.5555/3545946.3598758}

\bibitem{DBLP:conf/atal/VSC23a}
V., S.H.C., Singh, M.P., Chopra, A.K.: {Mandrake: Multiagent Systems as a Basis
  for Programming Fault-Tolerant Decentralized Applications}. In: Agmon, N.,
  An, B., Ricci, A., Yeoh, W. (eds.) Proceedings of the 2023 International
  Conference on Autonomous Agents and Multiagent Systems, {AAMAS} 2023, London,
  United Kingdom, 29 May 2023 - 2 June 2023. pp. 1218--1220. {ACM} (2023).
  \doi{10.5555/3545946.3598765},
  \url{https://dl.acm.org/doi/10.5555/3545946.3598765}

\bibitem{Winikoff2015}
Winikoff, M., Cranefield, S.: {On the testability of BDI agent systems}. IJCAI
  International Joint Conference on Artificial Intelligence pp. 4217--4221
  (2015)

\end{thebibliography}

\end{document}